\documentstyle{article}
\begin{document}
\input epsf
\rightline{UM-P-96/78}
\rightline{RCHEP 96/08}

\vskip 1.2cm
\begin{center}
\begin{large}
{\bf Baryogenesis from baryon number violating scalar interactions.}
\vskip 1.5cm
\end{large}
J. P. Bowes
and R. R. Volkas
\end{center}
\vskip 1cm
\noindent
Research Centre for High Energy Physics\\
School of Physics\\
University of Melbourne\\
Parkville, 3052\\
Australia.

\vskip 2cm
\begin{center}
{\bf Abstract}
\end{center}
\vskip 1cm
In the following work we consider the possibility of 
explaining the observed baryon number asymmetry in the universe from 
simple baryon number violating modifications, involving massive scalar 
bosons, to the Standard Model. 
In these cases baryon number violation is mediated through 
a combination of Yukawa and scalar self-coupling interactions.
Starting with a previously compiled catalogue of baryon-number violating 
extensions of the Standard Model, we identify the minimal subsets which 
can induce a $B-L$ asymmetry and thus be immune to sphaleron 
washout. For each of these models, we identify the region 
of parameter space that leads to the production of a baryon number 
asymmetry of the correct order of magnitude.

\newpage

\begin{center}
{\bf 1. Introduction and Motivation}
\end{center}

The excess of matter over antimatter observed in the present day universe 
can be explained by proposing that there existed suitable baryon number 
violating physical processes which operated in the early 
stages of the universe. 
These physical processes led to the situation whereby the production of 
baryons was slightly greater than 
the production of antibaryons thereby producing the required baryon-antibaryon 
asymmetry. In fact the quark-antiquark asymmetry required in the early 
universe ($t<10^{-6}$ sec) to give rise to the present excess of 
matter over antmatter
is very small:
\begin{equation}
{n_q-n_{\bar{q}} \over n_q}\simeq 3\times 10^{-8}.
\end{equation}
Thus the processes leading to the quark-antiquark asymmetry need only be 
minute to account for baryogenesis.

The three basic requirements for baryogenesis, which were first compiled in 
\cite{sakarov}, are: {\it(a) Baryon number violation}: if baryon number was
 not violated then the present day baryon asymmetry could only be 
explained by an initial baryon number asymmetry; {\it (b) C and CP violation}: 
even with the existence of baryon number violation, a baryon asymmetry 
would not have arisen unless there was a preference for the production of 
matter or antimatter in the early universe; and {\it (c) Non equilibrium 
conditions}: if the universe were in a continual state of thermal 
equilibrium, the phase space densities of baryons and antibaryons 
would neccessarily be equal, owing to the fact that by CPT 
invariance the baryon and antibaryon masses are always equal.

Ref\cite{bowes} provides a catalogue of all of the 
simplest extensions of the Standard Model that 
explicitly violate baryon number. (With lepton number 
broken explicitly through neutrino Majorana masses, all of the global $U(1)$ 
symmetries of the Standard Model are broken and an understanding of charge
quantisation results \cite{symm}. The charge quantisation 
issue provides a strong theoretical motivation for constructing 
all of the simplest Standard Model 
extensions which explicitly break $B$. See Ref\cite{bowes} for 
further discussion.) 
In each of these models baryon number violation is mediated through 
scalar bosons, which are required to be massive so as to 
ensure that nucleon decay remain unobserved. 
In the following work we will be considering the possibility that the 
out of equilibrium 
decays of these massive scalar bosons in the early universe could 
account for the present day baryon number asymmetry.
This work differs from similar calculations 
arising from GUT models \cite{weinberg} in that most of the models 
we will consider break baryon number 
through a combination of Yukawa interactions and scalar self-couplings
rather than through the baryon 
number violating Yukawa interactions 
associated with a single massive particle.

Starting with the
 baryon number violating interactions proposed in  \cite{bowes} we 
are left with the task of both demonstrating the existence of and calculating 
the magnitude of the CP violation arising in each of the baryon number 
violating models. 
In each case there will be two opposing decay paths for our massive 
scalar each giving rise to products with different baryon numbers. 
We can analyse this type of system using the following procedure:

If we assume for simplicity that our scalar boson $X$ has the two 
quark lepton final states $\bar{q}\bar{l}(B=-1/3)$ and $qq(B=2/3)$, then 
C and CP are violated if the branching ratio of the $X$ to the 
$\bar{q}\bar{l}$ final state $(=r)$ 
is unequal to the branching ratio of the $\bar{X}$ to the $ql$ 
final state $(=\bar{r})$, i.e. $r\neq\bar{r}$. We also know that CPT 
invariance requires that the total decay rates of $X$ and $\bar{X}$ be equal. 
Therefore we can write the branching ratios of the corresponding decays 
$X\rightarrow qq$ and $\bar{X}\rightarrow \bar{q}\bar{q}$, as $1-r$ 
and $1-\bar{r}$ respectively. From a symmetric initial state 
consisting of $X$ and $\bar{X}$, provided there are no further baryon number 
violating reactions, a net baryon asymmetry will exist after all of the 
$X$ and $\bar{X}$ decay, with the mean net baryon number produced 
being given by,
\begin{equation}
\Delta B=-\frac{1}{3}r+\frac{2}{3}(1-r)+\frac{1}{3}\bar{r}-
\frac{2}{3}(1-\bar{r})
=-(r-\bar{r}).
\end{equation}
We can therefore evaluate the mean net baryon number produced by calculating 
the difference in the branching ratio for boson and antiboson decay for just 
one of the decay channels.

We are hence essentially left with the task of calculating $r-\bar{r}$, which 
is a measure of the CP violation in the system. CP violating effects come from 
considering the interaction of the lowest order decay diagrams together with 
their loop corrections. It can be shown that the intermediate states in 
the loops must not only have CP violating complex couplings, but must also 
propagate on shell, thereby leading to complex Feynman amplitudes. Another 
important requirement is 
that the loop corrections involve baryon number violating interactions (see 
\cite{wolf}). In general 
we expect $\Delta B$ to be of order $\alpha^N$, where $\alpha$ 
characterises the coupling constants of the loop particles, and $N$ is the 
number of loops in the lowest order diagram which interferes with the 
tree graph to give a non zero value for $\Delta B$.\\

\begin{center}
{\bf 2. The Model:}
\end{center}

The following modifications to the standard model were devised with the 
primary aim of obtaining complete electric charge quantisation from 
the gauge invariance of the Lagrangian. 

Complete charge quantisation through gauge invariance can be obtained provided there is only one unembedded $U(1)$ invariance, and we obtain the correct 
charge quantisation provided the generator of this $U(1)$ invariance is 
the standard weak hypercharge $Y$. If there is more than one unembedded 
$U(1)$ invariance then the actual weak hypercharge of the theory can be chosen to be some linear combination of the standard model hypercharge and these 
additional symmetries of the theory.

As it stands the three-generation minimal standard model has five $U(1)$ 
invariances, the standard weak hypercharge $Y$, baryon number $B$, and the 
three family lepton numbers $L_e$, $L_{\mu}$, and $L_{\tau}$. These five $U(1)$ invariances correspond to there being four classically undetermined electric charges. To remove these invariances we must construct extensions of the minimal standard model which break $B$ and $L_i$ but leave $Y$ exact.

The simplest and most interesting way to break $U(1)_{L_i}$ is to introduce non-zero neutrino masses. This is most easily done by introducing right handed 
neutrinos into the model, where we choose that our right and left handed 
neutrinos be related through Dirac and Majorana mass terms.

This leaves us with just one undetermined electric charge, which can be taken 
to be the electric charge of the down quark $Q(d)$, or equivalently the hypercharge of the down quark $y_d$. Our four parameter 
uncertainty has thus been reduced to a two parameter uncertainty by this 
simple extension of the lepton sector. For more information see \cite{bowes}.

We are therefore left with the task of breaking this unwanted baryon 
symmetry without affecting the $U(1)_Y$ hypercharge symmetry.
This dual requirement is acheived by extending the standard model 
further to include either one or two exotic scalar multiplets 
together with a set of baryon number violating 
Yukawa and scalar-scalar self interactions.  
For further information on these 
models see \cite{bowes}.

The Yukawa interactions and the $SU(3)_C\otimes SU(2)_L\otimes U(1)_Y$ 
representations, where the hypercharge $Y$ is parameterised in terms of the 
unknown down quark hypercharge $y_d$, is listed below:
\begin{eqnarray}
\label{ferm}
\begin{array}{lclcl}
\sigma_{1.1}& \sim& \bar{Q}_L(f_L)^c\sim\bar{u}_R(e_R)^c
\sim\bar{d}_R(\nu_R)^c&
 \sim& (\bar{3},1,-y_d)(-1/3)\\
\sigma_{1.2}& \sim& \bar{Q}_L(f_L)^c& \sim& (\bar{3},3,-y_d)(-1/3)\\ 
\sigma_2& \sim& \bar{Q}_Le_R\sim\bar{u}_Rf_L& 
\sim& (\bar{3},2,-3-y_d)(-1/3)\\
\sigma_{3.1}& \sim& \bar{Q}_L(Q_L)^c\sim\bar{u}_R(d_R)^c &
 \sim& (3,1,-2-2y_d)(-2/3)\\ 
\sigma_{3.2}& \sim& \bar{Q}_L(Q_L)^c& \sim& (3,3,-2-2y_d)(-2/3)\\
\sigma_{3.3}& \sim& \bar{Q}_L(Q_L)^c\sim\bar{u}_R(d_R)^c &
 \sim& (\bar{6},1,-2-2y_d)(-2/3)\\
\sigma_{3.4}& \sim& \bar{Q}_L(Q_L)^c& \sim& (\bar{6},3,-2-2y_d)(-2/3)\\
\sigma_4& \sim& \bar{u}_R(\nu_R)^c& \sim& (\bar{3},1,-2-y_d)(-1/3)\\
\sigma_5& \sim& \bar{d}_Rf_L
\sim\bar{Q}_L\nu_R& \sim& (\bar{3},2,-1-y_d)(-1/3)\\
\sigma_{6.1}& \sim& \bar{u}_R(u_R)^c& \sim& (3,1,-4-2y_d)(-2/3)\\
\sigma_{6.2}& \sim& \bar{u}_R(u_R)^c& \sim& (\bar{6},1,-4-2y_d)(-2/3)\\
\sigma_{7.1}& \sim& \bar{d}_R(d_R)^c& \sim& (3,1,-2y_d)(-2/3)\\
\sigma_{7.2}& \sim& \bar{d}_R(d_R)^c& \sim& (\bar{6},1,-2y_d)(-2/3)\\
\sigma_{8}& \sim& \bar{d}_R(e_R)^c& \sim& (\bar{3},1,2-y_d)(-1/3).
\end{array}
\end{eqnarray}
Note that we have used the following notation for the standard model fermions
and right handed neutrinos:
\begin{eqnarray}
&f_L&\sim(1,2,-1),\;\;\;\;\;e_R\sim(1,1,-2),\;\;\;\;\;\nu_R\sim(1,1,0),
\nonumber\\
&Q_L&\sim(3,2,1+y_d),\;\;\;\;u_R\sim(3,1,2+y_d),\;\;\;\;
d_R\sim(3,1,y_d).
\end{eqnarray}
The fermion interactions, $\bar{Q}_L(Q_L)^c$, 
$\bar{Q}_L(Q_L)^c$, $\bar{u}_R(u_R)^c$, and $\bar{d}_R(d_R)^c$
 associated with the 
$\sigma_{3.2}$, $\sigma_{3.3}$, $\sigma_{6.1}$, and $\sigma_{7.1}$ scalars
are flavour antisymmetric. 

The simplest baryon number breaking scalar interactions were obtained by 
noting that certain pairs of scalars in the above list have group properties
which can be related by conjugation subject to the required constraint $y_d=-2/3$. 
For these cases the two distinct scalars listed in Eq(\ref{ferm}) can be 
considered as representing 
one particle with two sets of baryon number breaking Yukawa 
interactions. There are three such conjugate pairs which we list below:
\begin{eqnarray}
\label{conj}
\sigma_{1.1}&=&\sigma_{3.1}^c\sim(\bar{3},1,2/3)\nonumber\\
\sigma_{1.2}&=&\sigma_{3.2}^c\sim(\bar{3},3,2/3)\nonumber\\
\sigma_4&=&\sigma_{7.1}^c\sim(\bar{3},1,-4/3).
\end{eqnarray}

Baryon number violation is also obtainable by considering the scalar-scalar 
interactions of the scalars listed above. 
By considering every possible scalar combination in Eq(\ref{ferm}),
two lists of possible charge quantising scalar potentials were 
compiled, corresponding to 
$\Delta B=1$ baryon number violating processes and $\Delta
B=2$ baryon number violating processes respectively. 
The $\Delta B=1$ list
is shown below: 
\begin{eqnarray}
\label{6}
\sigma_1,\sigma_2&\rightarrow&\sigma_{1.2}\sigma_{1.2}\sigma_2\phi\nonumber\\
\sigma_1,\sigma_3&\rightarrow&
\sigma_{1.1}\sigma_{3.1}+\sigma_{1.1}\sigma_{1.1}^c\sigma_{1.1}\sigma_{3.1}+
\sigma_{1.1}\sigma_{3.1}\sigma_{3.1}^c\sigma_{3.1}+\sigma_{1.1}\sigma_{3.1}
\phi^{\dagger}\phi\nonumber\\
&\rightarrow&\sigma_{1.2}\sigma_{3.2}+\sigma_{1.2}\sigma_{1.2}^c\sigma_{1.2}
\sigma_{3.2}+ \sigma_{3.2}\sigma_{3.2}^c\sigma_{3.2}\sigma_{1.2}
+\sigma_{1.2}\sigma_{3.2}\phi^{\dagger}\phi \nonumber\\
\sigma_1,\sigma_5&\rightarrow&\sigma_{1.1}\sigma_5\sigma_5 \nonumber\\
&\rightarrow&\sigma_{1.2}\sigma_{1.2}\sigma_5\phi^c\nonumber \\
\sigma_1,\sigma_6&\rightarrow&\sigma_{1.2}\sigma_{6.1}\phi\phi\nonumber\\
\sigma_1,\sigma_7&\rightarrow&\sigma_{1.2}\sigma_{7.1}\phi^c\phi^c\nonumber\\
\sigma_2,\sigma_3&\rightarrow&\sigma_2^c\sigma_{3.2}\sigma_{3.2}\phi^c
\nonumber\\
\sigma_2,\sigma_7&\rightarrow&\sigma_2\sigma_{7.1}\phi\nonumber\\
\sigma_3,\sigma_4&\rightarrow&\sigma_{3.2}\sigma_4\phi\phi \nonumber\\
\sigma_3,\sigma_5&\rightarrow&\sigma_{3.1}\sigma_5\phi\nonumber\\
&\rightarrow&\sigma_{3.2}\sigma_{3.2}\sigma_5^c\phi\nonumber \\
\sigma_3,\sigma_{8}&\rightarrow&\sigma_{3.2}\sigma_{8}\phi^c\phi^c\nonumber\\
\sigma_4,\sigma_7&\rightarrow&\sigma_4\sigma_{7.1}+
\sigma_4\sigma_4^c\sigma_4\sigma_{7.1}+
\sigma_{7.1}\sigma_{7.1}^c\sigma_{7.1}\sigma_4+
\sigma_4\sigma_{7.1}\phi^{\dagger}\phi\nonumber\\
\sigma_5,\rho&\rightarrow&\sigma_5\sigma_5\sigma_5\rho\nonumber\\
\sigma_5^a,\sigma_5^b&\rightarrow
&\sigma_5^a\sigma_5^b\sigma_5^b\phi\nonumber\\
\sigma_5,\sigma_7&\rightarrow&\sigma_5\sigma_{7.1}\phi^c\nonumber\\
\sigma_6,\sigma_{8}&\rightarrow&\sigma_{6.1}\sigma_{8}
+\sigma_{6.1}\sigma_{6.1}^c\sigma_{6.1}\sigma_8+\sigma_8\sigma_8^c
\sigma_8\sigma_{6.1}+\sigma_{6.1}\sigma_8\phi^{\dagger}\phi
\end{eqnarray}
where $\phi$ represents the SM Higgs scalar $\phi\sim(1,2,1)$, and $\rho$
represents a new Higgs like scalar $\rho\sim(8,2,1)$.   
Similarly the $\Delta B=2$ list of scalar potentials take the form:
\begin{eqnarray}
\label{7}
\sigma_1,\sigma_3&\rightarrow&\sigma_{1.1}
\sigma_{3.1}\sigma_{1.1}\sigma_{3.1}\nonumber\\
&\rightarrow&\sigma_{1.1}\sigma_{3.2}\sigma_{1.1}\sigma_{3.2}\nonumber\\
&\rightarrow&\sigma_{1.1}\sigma_{3.3}\sigma_{1.1}\sigma_{3.3}\nonumber\\
&\rightarrow&\sigma_{1.1}\sigma_{3.4}\sigma_{1.1}\sigma_{3.4}\nonumber\\
&\rightarrow&\sigma_{1.2}\sigma_{3.1}\sigma_{1.2}\sigma_{3.1}\nonumber\\
&\rightarrow&\sigma_{1.2}\sigma_{3.2}\sigma_{1.2}\sigma_{3.2}\nonumber\\
&\rightarrow&\sigma_{1.2}\sigma_{3.3}\sigma_{1.2}\sigma_{3.3}\nonumber\\
&\rightarrow&\sigma_{1.2}\sigma_{3.4}\sigma_{1.2}\sigma_{3.4}\nonumber\\
\sigma_3,\sigma_7&\rightarrow&\sigma_{3.1}\sigma_{3.1}\sigma_{7.2}\nonumber\\
&\rightarrow&\sigma_{3.2}\sigma_{3.2}\sigma_{7.2}\nonumber\\
&\rightarrow&\sigma_{3.3}\sigma_{3.3}\sigma_{7.2}\nonumber\\
&\rightarrow&\sigma_{3.4}\sigma_{3.4}\sigma_{7.2}\nonumber\\
\sigma_4\sigma_7&\rightarrow&
\sigma_4\sigma_{7.1}\sigma_4\sigma_{7.1}\nonumber\\
&\rightarrow&\sigma_4\sigma_{7.2}\sigma_4\sigma_{7.2}\nonumber\\
\sigma_6,\sigma_7&\rightarrow&\sigma_{6.2}\sigma_{7.1}\sigma_{7.1}\nonumber\\
&\rightarrow&\sigma_{6.2}\sigma_{7.2}\sigma_{7.2}.
\end{eqnarray}

The above scalar potential terms can be placed into groups consisting of
quadratic, cubic and quartic terms. As shown in \cite{bowes} each of 
these different groups of scalar interactions are of different 
phenomenological 
interest, with the higher order interactions (i.e. the quartics and cubics) 
having the least stringent experimental constraints (obtained from 
nucleon decay data) on $m_{\sigma}$, the 
mass of the scalar. Each of these subgroups will 
also result in different outcomes as far as explaining baryogenesis is 
concerned. 
In the following work we will be determining which 
of the above interactions are able to account for baryogenesis.
This will basically involve the calculation of $r-\bar{r}$, the CP 
violation arising from each model and the subsequent use of standard 
cosmological arguments. 
\\

\begin{center}
{\bf 3. Calculation of Baryon Production:}
\end{center}

As previously mentioned 
baryogenesis is only possible provided we have CP violation 
incorporated in such a way as to induce different partial decay 
rates for baryon number violating decays of particles and antiparticles. 
To obtain this CP violation we require the introduction of at least two 
baryon number violating scalar self interactions. For example for the 
$\sigma_{1.1}\sigma_5\sigma_5$ class of scalar potentials
 which we will consider 
later, we are required to introduce two scalar interactions, 
$\sigma_{1.1}\sigma_5\sigma_5$ and $\sigma_{1.1}\sigma'_5\sigma'_5$, where the 
scalars $\sigma_5$ and $\sigma'_5$ 
have identical group properties but different masses and couplings. 
There are two reasons behind this requirement;
 the first being that if the 
exchanged scalar(s) $Y$ in the loop corrections have identical masses and 
properties to the decaying particle(s) X, then the contribution to 
baryogenesis arising from the decay of $X$ with the exchange of $Y$ 
will be cancelled by an equal 
and opposite contribution made by the decay of $Y$ with the exchange of $X$;
the second reason arises from the requirement that there be an 
imaginary component 
to the product of coupling constants associated with the CP violating 
tree and loop graphs, which is impossible if each Yukawa coupling appears 
with its conjugate.

The above list of baryon number violating 
models can be narrowed down if we consider the 
effects of sphaleron or other forms of damping processes 
on any baryon number asymmetry produced by each model \cite{kolbaturner}.
The rapid $B+L$ violating sphaleron transitions that are still occuring after 
the decays of our $\sigma$-particles will quickly erase any $B+L$ asymmetry.
If we take this factor into 
account we can rule out a number of the models listed in 
Eqs(\ref{conj},\ref{6},\ref{7}), by keeping only those that produce 
a $B-L$ asymmetry (these being immune from sphaleron washout)
 we are left with the 
following much shortened list of possible baryon asymmetry generating models:
\begin{eqnarray}
\label{reduced}
\sigma_1,\sigma_5&\rightarrow&\sigma_{1.1}\sigma_5\sigma_5 \nonumber\\
\sigma_2,\sigma_7&\rightarrow&\sigma_2\sigma_{7.1}\phi\nonumber\\
\sigma_3,\sigma_5&\rightarrow&\sigma_{3.1}\sigma_5\phi\nonumber\\
\sigma_5,\sigma_7&\rightarrow&\sigma_5\sigma_{7.1}\phi^c\nonumber\\
\end{eqnarray}
Thus if we allow for damping, 
we are left with just four interactions, one cubic, and three Higgs  
containing cubic interactions.

It should also be noted that the version of the SM which we have used 
includes Majorana masses for neutrinos, which can in themselves 
allow for the production of a baryon asymmetry or the damping of 
a preexisting baryon asymmetry. 
If $M>m_{\sigma}$, where $M$ is a Majorana neutrino mass, then 
out-of-equilibrium decays of $\nu_R$ may produce a $\Delta L$ prior 
to the decays of the $\sigma$ bosons. However, this asymmetry 
will be erased, and in particular it will {\it not} be reprocessed into 
a $\Delta B$ through $B+L$ violating sphaleron processes. This is 
because the $B-L$ violating $\sigma$ interactions (which we assume 
still occur rapidly) will combine with sphaleron processes to force 
both $\Delta L$ and $\Delta B$ to vanish. If $M<m_{\sigma}$, then 
any $\Delta B$ produced by $\sigma$ decays will be erased by the combination 
of rapid $L$ violating and $B+L$ violating processes. Subsequent 
out-of-equilibrium decays of $\nu_R$ may produce a $\Delta L$ which gets 
reprocessed into a $\Delta B$ by sphaleron effects, given that the 
$\sigma$-induced $B-L$ violating processes have already switched off. 
Although this is an interesting scenario 
(and has been studied in other contexts \cite{leptonexchange}), 
it is not the one we wish to consider in 
this paper. We will therefore require that $M>m_{\sigma}$, so that only 
$\sigma$-decays contribute to the $\Delta B$ that survives to the present 
day. (We also need to assume that no baryon asymmetry is produced during 
the electroweak phase transition. This, however, is assured provided 
that we assume the existence of only a single electroweak Higgs doublet 
$\phi$, see for example Ref\cite{higgs} for a review.)

Although any baryon number asymmetry arising from the conjugate 
pair and quadratic interactions will be damped away, 
we consider these systems first as an instructive exercise in how to do
the calcuations. The realistic cases will then follow by simple extension.

The conjugate pair interactions are the simplest class of interaction 
included in our catalogue. The calculation of the baryogenesis 
arising from these conjugate pair interactions is directly 
analogous to that obtained by \cite{weinberg} using a GUT model. 
CP violation arises from considering the tree and one loop 
corrections for the scalar decay \cite{weinberg}. For the 
$\sigma_{1.1}-\sigma^c_{3.1}$ system, our Yukawa terms have the following 
form,
\begin{eqnarray}
\cal{L}&=&\lambda_1(\bar{e}_R)^c\sigma u_R+
\lambda_3\bar{u}_R\sigma(d_R)^c \\ \nonumber
&+&\lambda'_1(\bar{e}_R)^c\sigma' u_R+\lambda'_3\bar{u}_R\sigma'(d_R)^c
+h.c. 
\end{eqnarray}
where $\sigma \sim\sigma'\sim(\bar{3},1,2/3)$ under $SU(3)_c\otimes
SU(2)_L\otimes U(1)_Y$. The reasons for introducing two copies of 
$(\bar{3},1,2/3)$ are discussed in the first paragraph of Sec 2 above.
This leads to the following expression for the baryon number production:
\begin{equation}
\label{10}
r-\bar{r}=
\frac{4 ImTr(\lambda_1^{\dagger} \lambda'_1 \lambda_3
{\lambda'_3} ^{\dagger})}
{Tr(\lambda_1 ^{\dagger}\lambda_1)+Tr(\lambda_3 ^{\dagger}\lambda_3)}
ImI(m_{\sigma '} /m_{\sigma}).
\end{equation}
The function $I(m_{\sigma'}/m_{\sigma})$ represents 
the Feynman integral for the exchange of the scalar $\sigma'$ 
in the decay of the scalar $\sigma$,
and it has the form
\begin{equation}
\label{11}
ImI(m_{\sigma '}/m_{\sigma})=-\frac{1}{16\pi}\left[1-\frac{m_{\sigma'}^2}
{m_{\sigma}^2}
\ln\left(1+\frac{m_{\sigma}^2}{m_{\sigma'}^2}\right)\right].
\end{equation}
If we assume that the mass of $\sigma$ is
 much larger than the mass of $\sigma'$ 
then we can ignore the contribution 
made by the decay of the $\sigma'$ scalar and Eq(\ref{10}) is 
essentially the complete 
contribution to baryogenesis.

In the quadratic models baryon number 
is violated through a scalar-scalar interaction. The 
analysis of the bilinear interaction coupling two different scalars 
is similar to that associated with 
 the conjugate pair case, and 
we must again introduce two sets of interactions to obtain  non zero CP 
violation. In our sample calculation we will use the 
$\sigma_{1.1}\sigma_{3.1}$ and $\sigma'_{1.1}\sigma'_{3.1}$ 
bilinears together with the following 
Yukawa interactions:
\begin{eqnarray}
\cal{L}&=&\lambda_1(\bar{e}_R)^c\sigma_1 u_R+
\lambda_3(\bar{d}_R)^c\sigma_3 u_R \\ \nonumber
&+&\lambda'_1(\bar{e}_R)^c\sigma'_1 u_R+\lambda'_3(\bar{d}_R)^c\sigma'_3 u_R
+h.c. 
\end{eqnarray}
The CP violating tree and one loop corrections for this model are shown in 
Fig(1). In this case we have a four vertex loop correction rather 
than the three vertex loop correction we had for the conjugate pairs. 
The CP violating Feynman amplitude corresponding to $\sigma_1$ decay 
takes the form:
\begin{equation}
\label{13}
M(\sigma_1 \rightarrow \bar{u}+\bar{e})=\lambda_1+\frac{(\mu^2)(\mu'^2)^
{\dagger}}{m_{\sigma_1}^2-m_{\sigma_3}^2}
\lambda'_1\lambda_3 ^{\dagger}\lambda'_3
I(m_{\sigma'_1}/m_{\sigma_1},m_{\sigma'_3}/m_{\sigma_1}),
\end{equation}
where $\mu^2$ is the coupling constant associated with the bilinear 
$\mu^2 \sigma_{1.1}\sigma_{3.1}$, and $\mu'^2$ is the coupling constant 
associated with the bilinear $\mu'^2\sigma'_{1.1}\sigma'_{3.1}$, 
where $\mu$ and $\mu'$ have dimensions of mass. The 
factor 
$I(m_{\sigma'_1}/m_{\sigma_1},m_{\sigma'_3}/m_{\sigma_1})$ again represents 
the contribution of the Feynman integral around the loop; 
in this case two scalars connected 
by a mass insertion are exchanged. 
By taking the difference between absolute value of Eq(\ref{13}) squared and 
the corresponding squared amplitude of the antiparticle process, we 
arrive at the following 
expression for the baryon number production:
\begin{equation}
\label{14}
r-\bar{r}=
\frac{4}{m_{\sigma_1} ^2(m_{\sigma_1} ^2-
m_{\sigma_3} ^2)}
\frac{ImTr\left ((\mu^2)(\mu'^2)^{\dagger}\lambda_1 ^{\dagger}\lambda'_1
\lambda_3 ^{\dagger}\lambda'_3 \right )}{Tr(\lambda_1 ^{\dagger}\lambda_1)}
ImI(m_{\sigma'_1}/m_{\sigma_1},m_{\sigma'_3}/m_{\sigma_1})
\end{equation}
The value of $I(m_{\sigma'_1}/m_{\sigma_1},m_{\sigma'_3}/m_{\sigma_1})$ in 
this case is found to have the form:
\begin{eqnarray}
\label{15}
ImI(\rho_1,\rho_3)&=&\frac{1}{32\pi}\frac{1}{\rho_1^2-\rho_3^2}(
-7\rho_1 ^2+7\rho_3 ^2+\rho_1 ^2\ln[1+\rho_1 ^{-2}]   \nonumber\\
&+&7\rho_1^4\ln[1+\rho_1^{-2}]
-\rho_3^2\ln[1+\rho_3^{-2}]-
7\rho_3^4\ln[1+\rho_3^{-2}])
\end{eqnarray}
where,
\begin{equation}
\rho_i=\frac{m_{\sigma'_i}}{m_{\sigma_1}}.
\end{equation}
Note the similarity between the two expressions for the imaginary components 
of the phases for the conjugate pair and quadratic cases, Eq(\ref{11})
 and Eq(\ref{15}), particularly in the limit
 $m_{\sigma'}/m_{\sigma} \rightarrow \infty$.
We assume that $m_{\sigma_3}>m_{\sigma_{1}}> m_{\sigma'_{1,3}}$ which 
as can be seen from Eqs(\ref{15}) and (\ref{14}) will 
allow us to ignore the contribution 
the decays of $\sigma'_{1,3}$ and $\sigma_3$ will make to baryogenesis.
Thus Eq(\ref{14}) can be regarded as constituting the entire 
contribution to baryogenesis. 

It should be noted that we can also obtain baryogenesis 
in the special case where we have just three new exotic scalars. 
For example for 
$\sigma_1$ decay we can set $\sigma_3=\sigma_3'$ and still 
obtain the required imaginary components to the Yukawa couplings and 
the Feynman integral.

We now turn to realistic models based on Eq(\ref{reduced}). In addition 
to not having a sphaleron washout problem, these models are also of 
greater phenomenological interest because the nucleon-decay bounds 
on $m_{\sigma}$ are much weaker than those for the unrealistic conjugate 
pair and quadratic toy models just considered ($m_{\sigma}>10^6$GeV 
for the $\sigma_{1.1}\sigma_5\sigma_5$ cubic compared with $m_{\sigma}>10^{15}$GeV for the conjugate pair and quartic interactions \cite{bowes}).

From a diagrammatic and calculational point of view these models differ 
from the above because baryon number violation arises from a cubic or 
trilinear term in the Higgs potential.

Consider first the $\sigma_{1.1}-\sigma_5$ system. We will initially 
introduce two 
copies of both the $\sigma_{1.1}$ and $\sigma_5$ scalars; we will 
require at least three of these four scalars to obtain 
the required CP violation. 
We will consider a simplified set of Yukawa 
interactions associated with the participating scalars as shown below:
\begin{eqnarray}
\label{tri}
\cal{L}&=&\lambda_1(\bar{e}_R)^c\sigma_1 u_R+\lambda_5
\bar{\nu}_L\sigma_{5a}d_R+\lambda_5\bar{e}_L\sigma_{5b}d_R\\ \nonumber
&+& \lambda'_1(\bar{e}_R)^c\sigma'_1 u_R+\lambda'_5
\bar{\nu}_L\sigma'_{5a}d_R+\lambda'_5\bar{e}_L\sigma'_{5b}d_R+h.c. 
\end{eqnarray}
where $\sigma_{5a}$ and $\sigma_{5b}$ designate the two $SU(2)$ components 
of $\sigma_5$. The above set of Yukawa interactions are simplified in that 
we have considered just a few of the 
Yukawa interactions, see Eq(\ref{ferm}), which may from 
group considerations be associated 
with the exotic scalars.
The CP violating tree and loop corrected diagrams
for the decay of the $\sigma_5$ multiplet
 are shown in Fig(2). For these decays the lowest order loop 
corrections involve both the $\sigma_1\sigma_5\sigma_5$ and 
$\sigma_1\sigma'_5\sigma'_5$ cubic interactions. (In the 
interests of simplicity we have ignored the contribution the 
$\sigma_1\sigma_5\sigma'_5$ interaction will make to baryogenesis.) 
Thus we require only 
one version of the $\sigma_1$ scalar to obtain baryogenesis from the 
decay of the $\sigma_5$ scalar, conversely we would require two versions 
of the 
$\sigma_1$ scalars and one $\sigma_5$ scalar to obtain baryogenesis 
from the decay of the $\sigma_1$ scalar.
The loop corrected diagrams shown in Fig(2) are complicated 
and as such won't be evaluated. 
This omission is justified because we are interested 
in an order of magnitude calculation only. In any case, it is the 
coupling constants 
which will play the major part in determining the numerical value 
of the baryogenesis arising from  
these cubic interactions.
Based on the results obtained for the
 conjugate pairs and the quadratic 
interactions it is reasonable to assume that the 
imaginary part of the Feynman integral will be of a similar form 
and hence give similar numerical 
values to those calculated in 
Eqs(\ref{11}) and 
(\ref{15}); we must however allow for an extra $1/(2\pi)^2$ suppression 
factor for each additional loop order. For the diagrams given in Fig(2) 
the imaginary component of our loop integral will thus be suppressed by 
 an extra factor of $1/(2\pi)^2$ in comparison to the 
one loop integrals considered in Eqs(\ref{11}) and 
(\ref{15}).
The CP violating amplitudes for the decays shown in Fig(2) are given by:
\begin{equation}
\label{19}
M(\sigma_5 \rightarrow \bar{d}+f)=\lambda_5+\mu\mu'^{\dagger}
\lambda'_5\lambda_5 ^{\dagger}\lambda'_5
I(m_{\sigma_1},m_{\sigma_5},m_{\sigma'_5}).
\end{equation}
where $\mu$ represents the coupling constant associated with the cubic 
interaction (again with units of mass), and 
$I(m_{\sigma_1},m_{\sigma_5},m_{\sigma'_5})$ represents 
the Feynman integral taken around 
the loops, which in this case will not be evaluated. From 
Eq(\ref{19}) we obtain the following 
expression for the total baryon number production arising from the 
decays of the $\sigma_{5a}$ and $\sigma_{5b}$ scalars, where 
$m_{\sigma_5}= m_{\sigma_{5a}}\simeq m_{\sigma_{5b}}$;
\begin{equation}
\label{20}
r-\bar{r}= \frac{8}{m_{\sigma_5}^2}
\frac{ImTr(\mu\mu'^{\dagger}\lambda_5 ^{\dagger}\lambda'_5
\lambda_5 ^{\dagger}\lambda'_5)}{Tr(\lambda_5 ^{\dagger}\lambda_5)}
ImI(m_{\sigma_1},m_{\sigma_5},m_{\sigma'_5}).
\end{equation}
Note that we have included an extra factor of two in the above expression 
to allow for the fact that we have two equal (as $m_{\sigma_{5a}}
\simeq m_{\sigma_{5b}}$) contributions to the baryon number production. 
If we assume that $m_{\sigma_5}>m_{\sigma'_5}$ then 
we can ignore the contribution 
the decay of $\sigma'_5$ will make to baryogenesis. Thus Eq(\ref{20}) is 
in effect the complete contribution the system given in Eq(\ref{tri}) will 
make to baryogenesis.

The phenomenologically less interesting 
Higgs containing trilinears ($m_{\sigma}>10^{11}$GeV \cite{bowes}), 
will lead to baryogenesis via CP violating loop diagrams such as 
that shown in 
Fig(3) for the $\sigma_{3.1}\sigma_5\phi$ cubic, where we have used the 
following Yukawa interactions,
\begin{eqnarray}
\label{higgyuk}
\cal{L}&=&\lambda_3(\bar{u}_R)^c\sigma_3 d_R+\lambda_5
\bar{\nu}_L\sigma_{5a}d_R\\ \nonumber
&+& \lambda'_3(\bar{u}_R)^c\sigma'_3 d_R+\lambda'_5
\bar{\nu}_L\sigma'_{5a}d_R+h.c. 
\end{eqnarray}
Note the similarity between the diagrams in Fig(2) and Fig(3).
For $\sigma_i$ decay resulting from 
the interactions 
$\mu\sigma_3\sigma_5\phi$ and $\mu'\sigma_3'\sigma_5'\phi$ we 
obtain an expression for $r-\bar{r}$ of the form,
\begin{equation}
\label{20d}
r-\bar{r}=\frac{4}{m_{\sigma_5}^2}
\frac{ImTr(\mu\mu'^{\dagger}\lambda_5^{\dagger} {\lambda_5}'
{\lambda_3'} ^{\dagger} \lambda_3)}
{Tr(\lambda_5^{\dagger} \lambda_5)}
ImI(m_{\sigma_5},m_{\sigma_3},m_{{\sigma_5}'},m_{{\sigma_3}'},m_{\phi}).
\end{equation}
If $\sigma_5$ is significantly more massive than $\sigma_3$, $\sigma'_3$ and 
$\sigma'_5$ then 
Eq(\ref{20d}) is essentially the complete contribution to baryogenesis.
In this case our expression for the Feynman loop integral 
will involve the 
Higgs scalar with $m_{\phi}\ll m_{\sigma}$.

Note that in the special case of $\sigma_3=\sigma'_3$ we can still get 
baryon number violation from $\sigma_5$ decays, and vice-versa for the 
decay of the $\sigma_3$ scalar. Thus as in the trilinear case the simplest 
possible baryon number violating system requires the introduction of 
two Higgs containing cubic interactions and three exotic scalars.  \\

\begin{center}
{\bf 4. Numerical estimates}
\end{center}

From our expressions for the baryon number creating $r-\bar{r}$ obtained in 
the previous section, we will attempt to numerically determine the 
likelihood of each model accounting for baryogenesis.

We know from \cite {weinberg2} that the decay of super heavy bosons 
and antibosons 
at temperatures sufficiently well below their masses will produce a cosmic 
baryon entropy ratio of the form;
\begin{equation}
\frac{kn_B}{s}=0.28\frac{N_X}{N}\Delta B,
\end{equation}
where $k$ is Boltzmann's constant, $N_X/N$ is the ratio of the number 
of boson helicity states to the number of light particle 
helicity states
($N_X/N \approx 10^{-2}$)
and $\Delta B$ represents the baryon number generation arising from the decay. 
Astronomical observations 
put $kn_B/s\approx 10^{-10}$ to $10^{-8}$, therefore 
$\Delta B$ must lie within the range $10^{-8}$ to $10^{-6}$, to account 
for the 
presently observed baryogenesis.

Our expression for the baryon number created by the cubic interactions 
Eq(\ref{20}) can be expressed as
\begin{equation}
\Delta B= \frac{8\mu\mu'^{\dagger}}{m_{\sigma_5}^2}\epsilon\lambda^2 
ImI(m_{\sigma_1},m_{\sigma_5},m_{\sigma'_5}).
\end{equation}
Similarly our expression for the baryogenesis arising from the Higgs 
containing cubics, see Eq(\ref{20d}), can be expressed as
\begin{equation}
\label{255}
\Delta B= \frac{4\mu\mu'^{\dagger}}{m_{\sigma_i}^2}\epsilon\lambda^2 
ImI(m_{\sigma_5},m_{\sigma_3},m_{\sigma'_5},m_{\sigma'_3},m_{\phi})
\end{equation} 
where $\lambda$ represents a typical value for the magnitude of 
the unknown Yukawa couplings 
in Eqs(\ref{20}) and (\ref{20d}), and 
$\epsilon$ is a phase angle characterising the average strength of the
CP violation associated with our scalar-scalar, and Yukawa interactions. 

In our analysis we will assume that the scalar self interacting and 
Yukawa couplings are approximately equal in magnitude. Using this 
assumption we will set out to find the region of parameter space 
within which these coupling constants must fall in 
order to account for baryogenesis.

The value of the imaginary component of the Feynman integral
for the above two cubic interactions is estimated 
from the corresponding expressions obtained from the conjugate pairs 
and quadratic interactions. After allowing for the extra $1/(2\pi)^2$ 
integration factor, it is thus assumed to fall within the range
$(2\pi)^{-2}(10^{-3}-10^{-2})$.
The value for the lower bound arises from the equal mass 
$m_{\sigma}=m_{\sigma'}$ and $\rho_i=1$ values of 
Eqs(\ref{11}), and (\ref{15}) respectively, and the upper 
bound arises from the consideration of the $m_{\sigma}\gg m_{\sigma'}$ 
limit of Eq(\ref{11}).

By taking all of these assumptions on board (for both classes of cubic 
interaction), it is found that the coupling 
constants $\lambda$ and $\mu$ must lie within the following range
to account for baryogenesis,
\begin{equation}
\lambda \epsilon^{1/4}\approx\frac{|\mu|}{m_{\sigma}} 
\epsilon^{1/4}\approx 10^{-1}-10^{-2}.
\end{equation}
If we assume for example that $1>\epsilon >10^{-3}$ rads, then 
the couplings
$\lambda$ and $|\mu|/m_{\sigma}$ must be of order $1-10^{-2}$ to 
account for baryogenesis.
\\

\begin{center}
{\bf 4. Conclusion:}
\end{center}

We have thus demonstated that baryogenesis can be induced by 
each of the model 
types listed in Eq(\ref{reduced}). 

In the numerical analysis of the various interactions listed in 
Eq(\ref{reduced}), it was found that both 
classes of cubic interactions can successfully account for 
baryogenesis provided the coupling constants fall within the 
range,
$\lambda\epsilon^{1/4}=\mu\epsilon^{1/4}/m_{\sigma}
\approx 10^{-1}-10^{-2}$. 
Given that 
the known Higgs Yukawa couplings range up to order unity for the top 
quark, it is not at all unreasonable that the unknown Yukawa's may fall 
within this parameter space.

We have thus 
demonstrated that at least some of the baryon number violating 
models introduced in \cite{bowes}, with 
the explicit aim of  
obtaining complete charge quantisation from gauge invariance will also 
allow for baryogenesis. It should however be noted that whilst 
we only need one of the baryon number violating scalar interactions listed in 
Eqs(\ref{conj},\ref{6},\ref{7}) 
(and consequently two exotic scalar multiplets), 
to obtain charge 
quantisation, we require the introduction of at least two scalar-scalar 
interactions 
and at least three exotic scalars to account for baryogenesis.
Thus the minimal requirement for charge quantisation 
is not sufficient to also give us baryogenesis.

\newpage
\begin{center}
{\bf Figure Captions:}
\end{center}
\vskip 10mm

Figure 1:\\ \\ 
CP violating tree and one loop corrected diagrams resulting from the decay of $\sigma_{1.1}$ in the $\sigma_{1.1}\sigma_{3.1}$ quadratic model.
\vskip 10mm
Figure 2:\\ \\
CP violating tree and loop corrected diagrams resulting from the decay of $\sigma_{5}$ in the $\sigma_{1.1}\sigma_5\sigma_5$ cubic model.
\vskip 10mm
Figure 3:\\ \\
CP violating tree and loop corrected diagrams resulting from the decay of $\sigma_{5}$ in the $\sigma_{3.1}\sigma_5\phi$ cubic model.

\newpage
\vskip 5mm
\centerline{\epsfbox{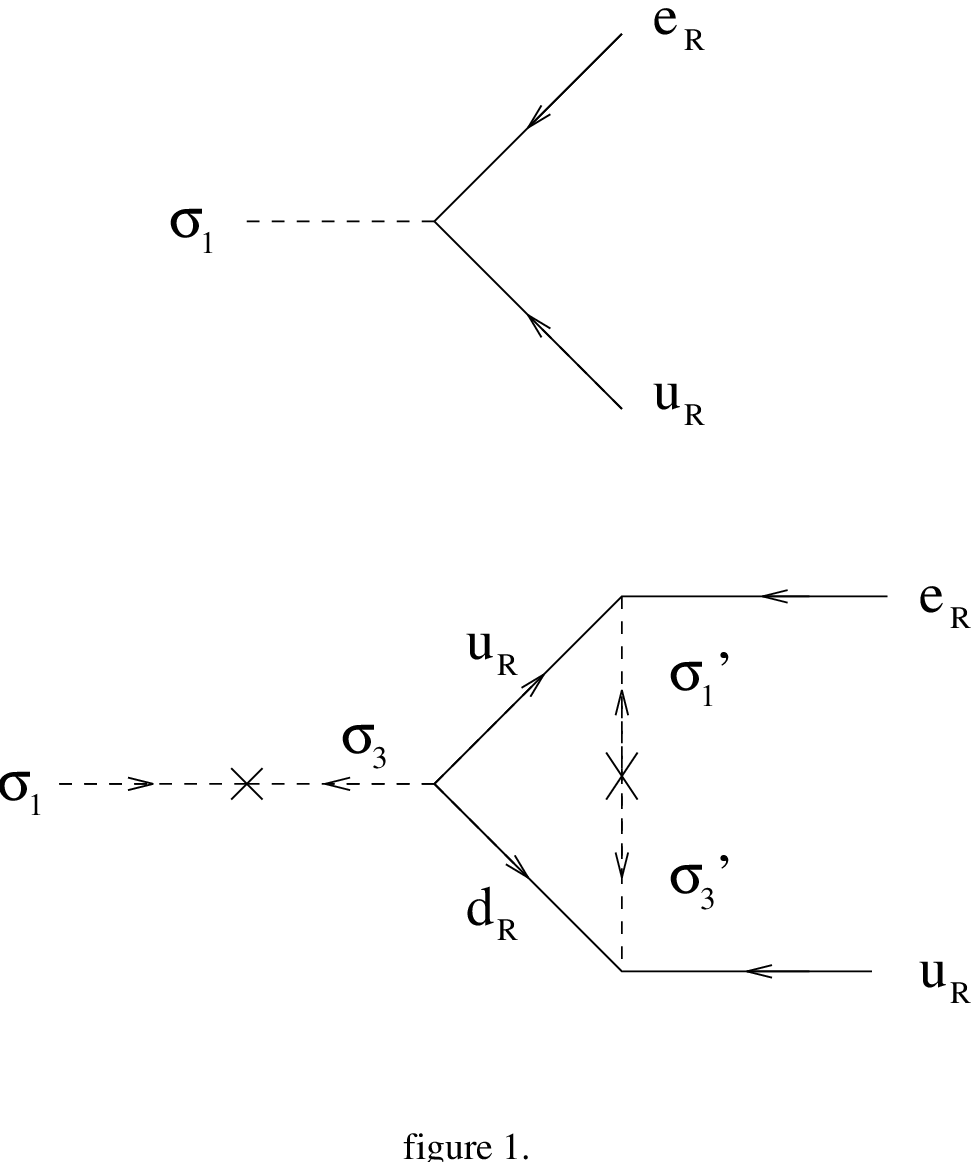}}
\vskip 50mm
\centerline{\epsfbox{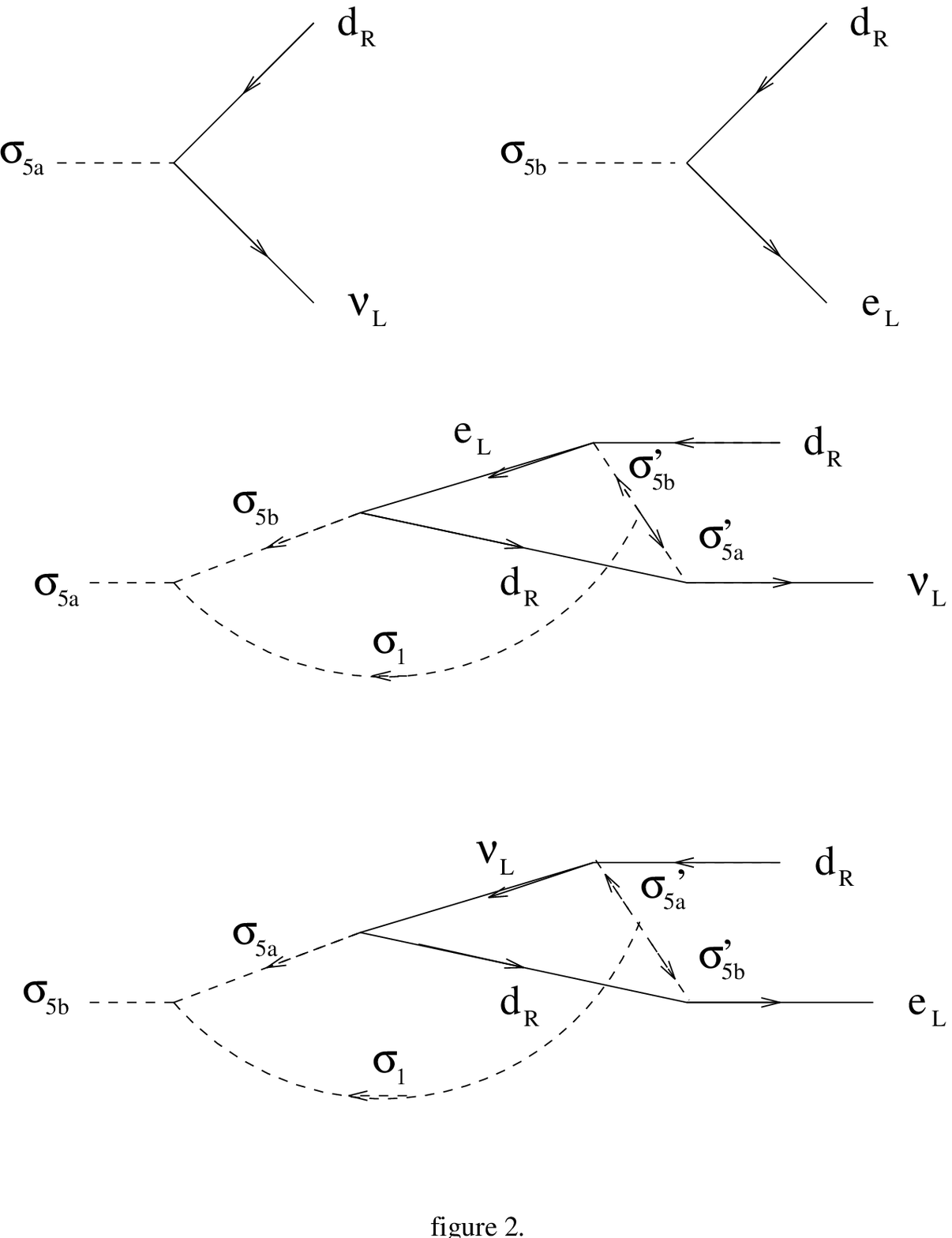}}
\centerline{\epsfbox{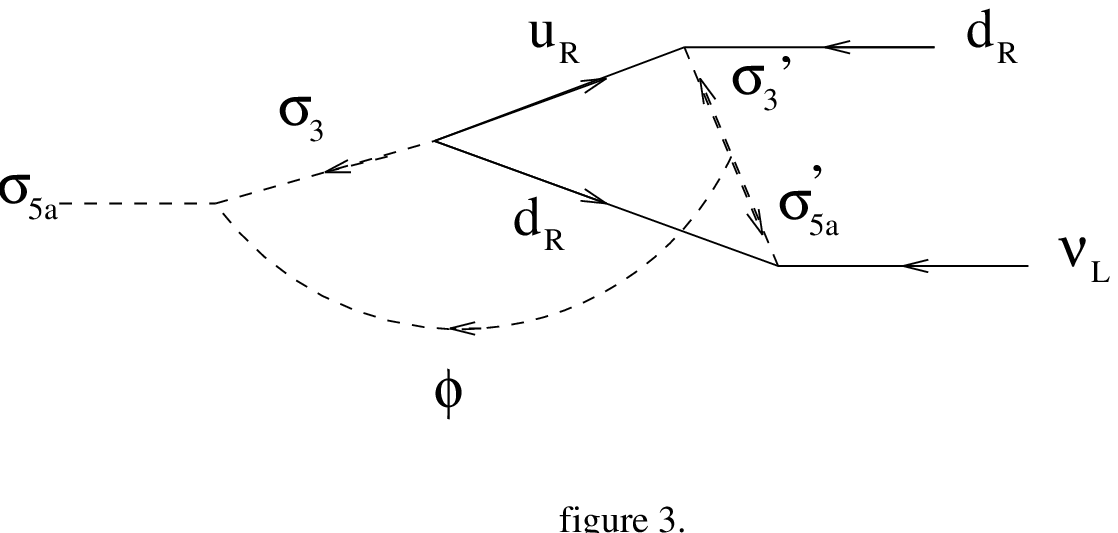}}
\vskip 50mm

\end{document}